\documentclass[10pt]{iopart}

\usepackage{iopams}

\usepackage{graphicx,xcolor}
\usepackage{booktabs}
\usepackage{amssymb}
\usepackage{hyperref}
\usepackage{comment}
\usepackage{booktabs}

\newcommand{\be}{\begin{equation}}
\newcommand{\ee}{\end{equation}}
\newcommand{\beq}{\begin{eqnarray}}
\newcommand{\eeq}{\end{eqnarray}}

\makeatletter
\usepackage{etoolbox}
\patchcmd{\@makefntext}{\fnsymbol}{\arabic}{}{}%
\patchcmd{\@thefnmark}{\fnsymbol}{\arabic}{}{}%
\def\@makefnmark{\textsuperscript{\arabic{footnote}}}%
\makeatother

\begin{document}

\title[Consistency of LIV neutrino scenarios in time delay analyses]{Consistency of Lorentz-invariance violation neutrino scenarios in time delay analyses}
\author{J.M. Carmona$^{a,b}$, J.L. Cort\'es$^{a,b}$, M.A. Reyes$^{a,b}$}

\address{$^a$ Departamento de F\'{\i}sica Te\'orica,
Universidad de Zaragoza, Zaragoza 50009, Spain}
\address{$^b$ Centro de Astropart\'{\i}culas y F\'{\i}sica de Altas Energ\'{\i}as (CAPA),
Universidad de Zaragoza, Zaragoza 50009, Spain}

\ead{jcarmona@unizar.es, cortes@unizar.es, mreyes@unizar.es}
\vspace{10pt}
\begin{indented}
\item[]\today
\end{indented}

\begin{abstract}
Modifications inspired by quantum gravity in  the kinematics of special relativity can manifest in various ways, including anomalies in the time of flight of massless particles and the emergence of decay channels for otherwise stable particles. Typically, these effects are studied independently; however, it may be necessary to combine both to perform a consistent analysis. In this work, we study the interplay between time-of-flight anomalies and neutrino instability in the context of a flavor-independent high-energy Lorentz-invariance violation (LIV) in the neutrino sector. Ensuring compatibility between both types of effects imposes strong constraints on the existence of early neutrinos with energies exceeding a maximum value determined by the scale of new physics. Such constraints depend on the specific LIV scenario and should be integrated into searches for high-energy neutrinos from gamma-ray bursts exhibiting LIV time shifts.
\end{abstract}

%
%
%
%
%

\section{Introduction}

The search for a quantum gravity theory motivates the study of departures from special relativity (SR) that become more significant when increasing the energy. The possible effect of new physics that we are going to discuss in this work is a Lorentz-invariance violation (LIV)~\cite{Colladay:1998fq}, motivated by the transition from the classical notion of spacetime to a quantum description of it.

The best candidates to look for these departures from SR are the astrophysical messengers, where we encounter energies orders of magnitude greater than those achievable with Earth-based accelerators. However, the use of observations of these messengers to reveal new physics is limited by uncertainties in their production mechanisms and in the description of the backgrounds through which they propagate.

Among the cosmic messengers, very high-energy astrophysical neutrinos stand out as one of the most promising candidates in the search for new physics. In the conventional scenario, their propagation is relatively simple to describe due to minimal interactions with backgrounds. However, when LIV is considered in the neutrino sector, unexpected effects during their propagation can lead to anomalies in the observation of neutrinos at Earth (see~\cite{Carmona:2022lyg} for a more extensive review).

A significant consequence of a violation of Lorentz invariance is the modification of the energy-momentum relation for free relativistic particles. This results in two distinct phenomenological effects in the neutrino sector. Firstly, the change in the relation between energy and momentum introduces energy-dependent neutrino propagation velocities, even in the massless limit. Consequently, this leads to a modification of their time of flight~\cite{Ellis:2002in,Jacob:2008bw}, which becomes observable when the distance between neutrino production and detection is sufficiently large, especially when the neutrino source is a rapidly transient event with minimal production timing uncertainty. This idea has been used in recent attempts to try to identify GRBs as potential sources of the highest energy neutrinos through directional and time correlations, modified by LIV (see \cite{ANTARES:2016fmg, Zhang:2022svg, Huang:2022xto, Amelino-Camelia:2022pja} for some examples).

Secondly, a modification of the neutrino dispersion relation has a profound impact on neutrino kinematics. This effect is comparable to the crucial role played by the mass, even if its contribution to the energy is negligible in the ultra-relativistic limit.  If the modification of the energy-momentum relation causes neutrinos to become superluminal, it results in neutrino instability and allows for new decay channels that were previously forbidden by SR kinematics~\cite{Carmona:2022dtp,Carmona:2012tp}. This drastic alteration influences then neutrino propagation and leads to observable changes in the spectrum detected on Earth (see~\cite{Mattingly:2009jf,Stecker:2014oxa,Reyes:2023osq} for some studies in this regard).

The two effects mentioned above are typically studied separately, as they provide alternative ways to put constraints on the magnitude of the deviation with respect to SR. However, we show in this work that, in order to perform a consistent analysis of neutrino time-of-flight anomalies, one should take into account the possibility of decays in the propagation of neutrinos due to the modified kinematics. The structure of the paper is the following. In Sec.~\ref{LIV} we describe a generic, flavour-independent, LIV correction in the neutrino sector and the modified dispersion relation that it produces.  
The corresponding change in the time of flight and the possible decays in the propagation to Earth are studied in Sec.~\ref{timedelay} and in Sec.~\ref{decay}, respectively. In Sec.~\ref{comparison}, we show how these two effects interplay in the case of astrophysical neutrinos, considering relevant neutrino energies and the magnitude of the LIV correction. To complement the analysis, we also study the non-standard neutrino propagation in the case of atmospheric neutrinos in Sec.~\ref{local}. Finally, in Sec.~\ref{conclusions} we summarize the most relevant conclusions.

\section{LIV for neutrinos and modified kinematics}
\label{LIV}

In order to introduce an LIV in the neutrino sector within the effective field theory framework, we need to write the possible additional LIV terms, depending on the neutrino fields, that can be added to the particle physics standard model (SM) Lagrangian. To implement effects that increase with the energy, one has to add LIV terms in the Lagrangian with dimension greater than four. We do not consider flavour-dependent terms, which are highly constrained from neutrino oscillations (see reviews~\cite{Mattingly:2005re,Stecker:2022tzd}). Looking at the free (quadratic in fields) part of the Lagrangian, one has a dominant term\footnote{Spatial derivatives do not appear in Eq.~(\ref{eq:LIV-nu}) because, treating LIV and masses as a perturbation, if we just consider the effects at first order, one  can use the field equations from the SM Lagrangian to replace space derivatives of fields by time derivatives in the additional LIV terms.}
\be
    \mathcal{L}_{\mathrm{LIV}}= - \frac{1}{\Lambda^n}\bar{\nu}_{L} \gamma^0(i\partial_0)^{n+1}\nu_{L} \,,
    \label{eq:LIV-nu}
\ee
where $\Lambda$ is the high-energy, `quantum-gravity', scale that parametrises the LIV. As a consequence of this additional LIV term in the free Lagrangian, the modified energy-momentum relation for neutrinos is
\be
E \,=\, |\vec{p}| \left[1+ \left(\frac{|\vec p|}{\Lambda}\right)^n\right]\,,
\ee
and for antineutrinos
\be
    E \,=\, |\vec{p}| \left[1+ (-1)^n \left(\frac{|\vec p|}{\Lambda}\right)^n\right]\,.
\ee
In the linear case ($n=1$) one has superluminal neutrinos and subluminal antineutrinos, while in the quadratic case ($n=2$) both neutrinos and antineutrinos are superluminal. 

Together with the previous two models, one has two additional models where the coefficient of the LIV term in the Lagrangian is chosen to be positive instead of negative. This produces a change of sign in the LIV correction to the expression of the energy as a function of the momentum and then an exchange of superluminal (subluminal) by subluminal (superluminal) particles. 

In addition to the LIV terms that are quadratic in fields, one could also introduce such terms in the interaction part of the SM Lagrangian. However, their effects would be negligible, as they would result in dynamical corrections proportional to $(E/\Lambda)$, with $(E/\Lambda)\ll 1$ for the energies accessible to us. This line of reasoning, however, does not apply to the LIV terms in the free part of the Lagrangian. These terms alter the dispersion relation and subsequently impact the kinematics. As indicated in the Introduction, such modifications can amplify subtle corrections, as is the case with reaction thresholds or the propagation over long distances. 


\section{Time of flight formula in LIV}
\label{timedelay}

For massless neutrinos, a modification of the neutrino energy-momentum relation produces an energy dependent velocity of propagation, 
\be 
    v= \frac{dE}{dp} \approx 1 \pm (n+1)\left( \frac{E}{\Lambda} \right)^{n} \,.
\ee
Therefore, the modification of the time of flight of a particle emitted by a source at a redshift $z$, due to a modification proportional to $(1/\Lambda)^n$ in the energy-momentum relation, is~\cite{Jacob:2008bw}\footnote{\label{DiffLambdas}Let us note that $\Lambda$ can be redefined up to a constant, and other works may use a different convention. For instance, in~\cite{Jacob:2008bw}, $\xi E_\mathrm{pl} \equiv \Lambda/2^{1/n}$, introducing a factor of $(1/2)$ in the expression of the time delay.}
\be
    \delta t_{\mathrm{LIV}}^{(n)} \,\approx\, \pm (n+1) \left(\frac{E}{\Lambda}\right)^n \frac{I_n(z)}{H_0}\,,
    \label{delta}
\ee
where $H_0=67.4 \,\mathrm{km\,s^{-1}\, Mpc^{-1}}$ and
\be 
    I_n(z)=\int_0^z dz' \frac{(1+z')^n}{\sqrt{\Omega_m (1+z')^3 + \Omega_\Lambda}} \,,
\ee
with $\Omega_m=0.315$ and $\Omega_\Lambda=0.685$\footnote{The numerical values of the constants that appear in this work are taken from \cite{Workman:2022ynf}.}. The $(+)$ and $(-)$ signs correspond to a subluminal and superluminal particle, respectively.

\begin{figure}[tbp]
    \centering
    \includegraphics[height=6cm]{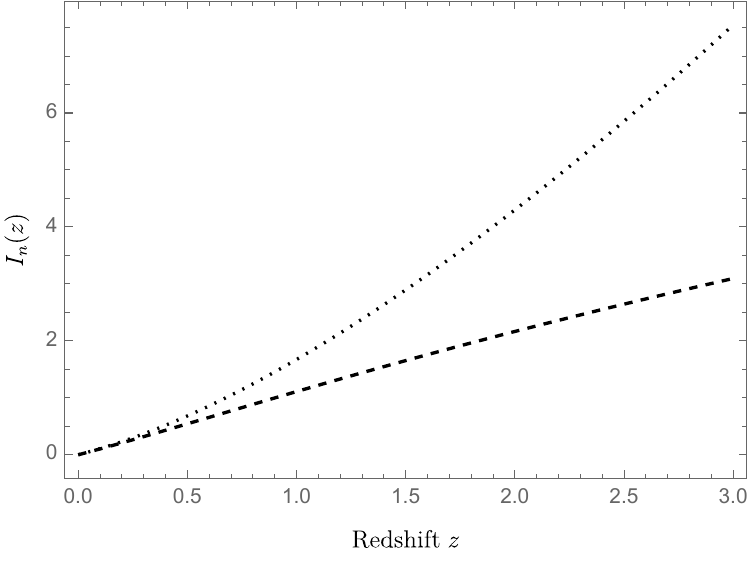}
    \caption{$I_n(z)$ for $n=1$ (dashed) and $n=2$ (dotted), as a function of the redshift of the source.}
    \label{fig:In}
\end{figure}

In this work we will focus on values of the scale of new physics close to the Plank scale ($M_P\approx 1.22 \cdot 10^{16}\,\mathrm{TeV}$), and astrophysical neutrinos with energies above $100\,\mathrm{TeV}$. Then, one obtains
\be
    \delta t_{\mathrm{LIV}}^{(1)} \approx \pm \,7.51\times 10^{3}\,\mathrm{s} \; \left(\frac{E}{100\mathrm{TeV}}\right) \left(\frac{\Lambda}{M_P}\right)^{-1} I_1(z)\,,
    \label{delta1}
\ee
for $n=1$, and 
\be
    \delta t_{\mathrm{LIV}}^{(2)} \approx \pm \, 9.23\times 10^{-11}\,\mathrm{s} \; \left(\frac{E}{100\mathrm{TeV}}\right)^2 \left(\frac{\Lambda}{M_P}\right)^{-2} I_2(z)\,,
    \label{delta2}
\ee
for $n=2$. The values of the dimensionless functions $I_1(z)$ and $I_2(z)$ for the redshifts of interest are shown in Fig.~\ref{fig:In}.

\section{Maximal energy due to LIV decays of superluminal neutrinos or antineutrinos}
\label{decay}

The modification of the energy-momentum relation due to LIV has another consequence that one has to consider together with the modification of the time of flight. When the neutrino or the antineutrino is superluminal, it becomes an unstable particle able to decay, producing either a neutrino-antineutrino pair or, above a certain energy threshold, also an electron-positron pair. These decays are a mechanism of loss of energy in the propagation of the superluminal particles that in the first case is accompanied by the production of a cascade of neutrinos and antineutrinos. 

\begin{figure}[tbp]
    \centering
    \includegraphics[width=\textwidth]{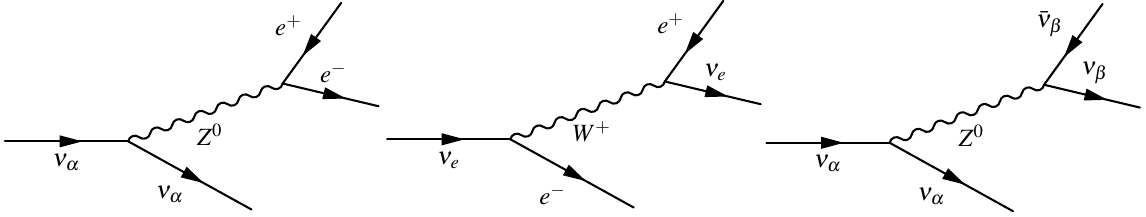}
    \caption{From left to right: (a) Neutral channel of the VPE; (b) Charged channel of the VPE; (c) Neutral channel of the NSpl.}
    \label{fig:feynman}
\end{figure}

The decay widths of electron-positron and neutrino-antineutrino production can be calculated in the collinear approximation~\cite{Carmona:2022dtp}. The diagrams one has to consider in the calculation  of the amplitude are those shown in Fig.~\ref{fig:feynman},
and the result for the decay width can be written in all the cases as
\be
    \Gamma(E) = 10^{-4}\, \kappa_n \,G_F^2 \,\frac{{E}^{5+3n}}{\Lambda^{3n}}\,,
    \label{eq:decaywidth}
\ee
where $G_F$ is the Fermi coupling constant, $G_F=1.17\times 10^{-5}\,\mathrm{GeV}^{-2}$, and the dimensionless constant $\kappa_n$ takes different values for the different decays (see Table~\ref{tab:kn}) due to the different couplings in the interactions involved and also due to the difference in the energy distributions (see~\cite{ReyesHung:2023udr} for more details).
\begin{table}[tbp]
\centering
\begin{tabular}{ccc} \toprule
    Decay & $\kappa_1$ & $\kappa_2$\\ \midrule
    $\nu_{\mu,\tau} \to \nu_{\mu,\tau} e^+ e^-$ & $1.01$ & $1.24$ \\
    $\nu_e \to \nu_e e^+ e^-$ & $13.0$ & $16.1$ \\ 
    $\nu_\alpha \to \nu_\alpha \nu \bar{\nu}$ & $3.87$ & $3.87$\\
    \bottomrule
\end{tabular}
\caption{Characteristic values of the constant $\kappa_n$ for the different decays and for $n=1$ and 2. For the case of the neutrino-antineutrino emission, $\kappa_n$ includes a factor 3 from the sum over the final neutrino flavours. The subindex $\alpha$ stands for the three different neutrino flavors: $e$, $\mu$ and $\tau$.}
\label{tab:kn}
\end{table}
It is important to remark that, according to Eq.~(\ref{eq:decaywidth}), all three decay widths are proportional to a high power of the energy ($8$ in the linear case and $11$ in the quadratic case). 

In the case of the decays producing an electron-positron pair, there is a restriction (kinematic threshold)
\be
    E > E_\mathrm{th}^{(n)} \doteq \left(2 m_e^2 \Lambda^n\right)^{1/(n+2)}\,,
    \label{Eth}
\ee
on the energy of the superluminal particle in order to be able to produce the pair of massive particles. In contrast, the neutrino-antineutrino pair production does not have any kinematical threshold in the approximation where we neglect neutrino masses.

When one considers the propagation of a superluminal particle in an expanding universe, one can determine the mean redshift interval for an electron-positron or neutrino-antineutrino decay from the ratio $\Gamma(E)/H_0$. If one defines an energy $E_r^{(n)}$ as the solution of the equation 
\be
    \Gamma(E)|_{E=E_r^{(n)}} \,=\, H_0\,,
    \label{Er(z)}
\ee
then, as a consequence of the strong energy dependence of the decay width, superluminal particles produced by a source at redshift $z$ with energy $E> E_r^{(n)}$ will decay in a very small $z$-interval, while those particles produced with an energy $E<E_r^{(n)}$ would need a very large $z$-interval to decay, so that they will arrive to the detector without decaying. The previous argument can be applied to the subsequent superluminal particles produced in the propagation of the superluminal particle emitted by the source at $z$. As a result, one will have a cascade of superluminal particles with energies close to $E_r^{(n)}$ in a nearby region (in terms of redshift) to the source that, together with the subluminal particles produced in the decays, will propagate as free particles with their energy being affected by the redshift due to the expansion of the universe. As a result of these arguments, one can estimate that all the superluminal particles originating from a source at $z$ will be detected with an energy $E_d$,
\be
    E_d < E^{(n)}_\mathrm{max}(z) \approx E_r^{(n)}/(1+z)\,,
\ee
i.e., one has a cutoff in the energy spectrum of superluminal particles due to LIV. 

In the decays producing an electron-positron pair, one also has to consider the kinematic threshold, Eq.~(\ref{Eth}). In fact, if electron-positron pair production were the only process at play, the maximum detected energy would be determined by the greater of the two energies, $E_r^{(n)}$  or $E_\mathrm{th}^{(n)}$, divided by $(1+z)$. In practise, the cutoff energy will be the lower of two values: the previous electron-positron cutoff energy  and the energy $E_r^{(n)}$ corresponding to the neutrino-antineutrino pair emission.

If we write the scale of new physics in units of the Plank scale, $E_r^{(n)}$ takes the values
\be 
    E_r^{(1)}\approx 4.58 \,\mathrm{TeV} \left( \frac{\Lambda}{M_P} \right)^{3/8}  \kappa_1^{-1/8}
    \label{eq:er1astro}
\ee
for $n=1$, and 
\be 
    E_r^{(2)}\approx 7.37\times 10^{4} \,\mathrm{TeV} \left( \frac{\Lambda}{M_P} \right)^{6/11} \kappa_2^{-1/11}
    \label{eq:er2astro}
\ee
for $n=2$.

Similarly, for $E_\mathrm{th}^{(n)}$, one obtains the following values:
\be 
    E_\mathrm{th}^{(1)} \approx 18.5 \,\mathrm{TeV} \; \left( \frac{\Lambda}{M_P} \right)^{1/3}
\ee
for $n=1$, and 
\be 
    E_\mathrm{th}^{(2)} \approx 9.39 \times 10^{4} \,\mathrm{TeV} \; \left( \frac{\Lambda}{M_P} \right)^{1/2}
\ee
for $n=2$.

\begin{figure}[tbp]
    \centering
    \includegraphics[height=6cm]{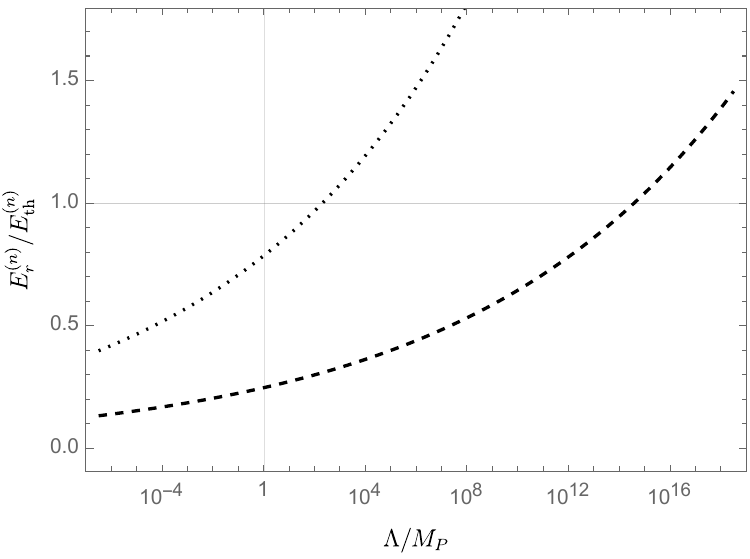}
    \caption{Quotient $E_r^{(n)}/E_\mathrm{th}^{(n)}$ as a function of $\Lambda/M_P$ for $n=1$ (dashed) and $n=2$ (dotted), under the approximation $\kappa_n^{-1/(5+3n)}\approx 1$.}
    \label{fig:ErEth}
\end{figure}

The different dependence on the LIV scale $\Lambda$ of the two energy scales, $E_r^{(n)}$ and $E_\mathrm{th}^{(n)}$, can be used to show that they become comparable when $\Lambda$ is larger than the Planck scale $M_P$ (see Fig.~\ref{fig:ErEth}).

Excluding this range of values for the scale of LIV,  the dominant decay of superluminal particles will be the decay producing a neutrino-antineutrino pair, and the corresponding maximum detected energy will be 
\be
   E_\mathrm{max}^{(1)}(z) \approx \frac{3.87 \,\mathrm{TeV}}{(1+z)} \left( \frac{\Lambda}{M_P} \right)^{3/8}  
   \label{Emax1}
\ee
for $n=1$, and
\be 
    E_\mathrm{max}^{(2)}(z) \approx \frac{6.52\times 10^{4} \,\mathrm{TeV}}{(1+z)} \left( \frac{\Lambda}{M_P} \right)^{6/11}
    \label{Emax2}
\ee
for $n=2$.

\section{Effects on the cosmological propagation of neutrinos and antineutrinos}
\label{comparison}

The effect of LIV in the time of flight of neutrinos has been used recently in order to see whether one could identify GRBs as the sources of the observed astrophysical high-energy neutrinos with an enlargement of the correlation time window between neutrino events and the low-energy gamma-ray observation of a GRB. In Refs.~\cite{Zhang:2022svg,Huang:2022xto}, several candidates of early and late events were seen to be compatible with a linear ($n=1$) modification of the energy-momentum relation of neutrinos and antineutrinos with\footnote{Refs.~\cite{Zhang:2022svg,Huang:2022xto} use a definition of the quantum-gravity scale that differs from ours in a factor of 2, see footnote~\ref{DiffLambdas}.} $(\Lambda/M_P)\sim(1/10)$. Using Eq.~(\ref{delta1}), the time delay for this LIV scale is
\be
    \delta t_{\mathrm{LIV}}^{(1)} \approx \pm 7.51\times 10^{4}\;\mathrm{s} \; \left(\frac{E}{100\,\mathrm{TeV}}\right) I_1(z)\,.
\ee
A similar result has been obtained in \cite{Amelino-Camelia:2022pja}, but in this case the use of the latest results of the IceCube experiment has changed the identification of possible GRB-neutrinos, finding only subluminal candidates with respect to an analysis based on earlier data that contained both early and late events~\cite{Amelino-Camelia:2016ohi}. In any case, no definite conclusion can be extracted at present from those analyses owing to the limited statistics and the uncertainties in the energy and the direction of neutrino events, as well as in the redshift assignments to the GRBs. 

All the attempts to use LIV to identify GRB-neutrinos have been based on the modification of the time of flight. However, according to Eq.~(\ref{Emax1}), an LIV scale $(\Lambda/M_P)\sim(1/10)$ for the linear case leads to a maximum energy of early events
\be
   E_\mathrm{max}^{(1)}(z) \approx 1.63 \;\mathrm{TeV} / (1+z) \,,
\ee
which is below the energy range of the observed astrophysical high-energy neutrino events.
Then, we conclude that no early events should be observed, a result compatible only with the results of one of the analyses. A confirmation of the absence of early events in future analyses based on more complete data would be an argument in favor of the presence of LIV effects in the propagation of neutrinos from GRBs. 

\begin{figure}[tbp]
    \centering
    \includegraphics[height=6cm]{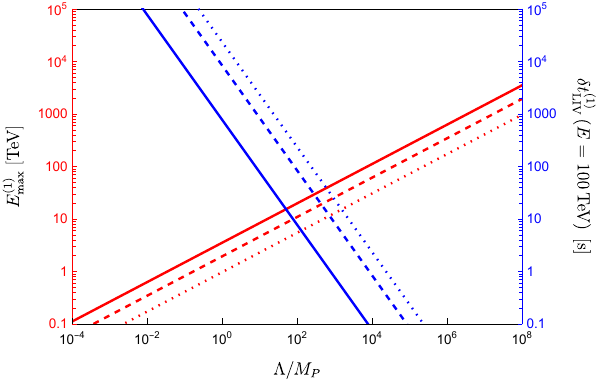}
    \caption{$E_\mathrm{max}^{(1)}$ of the neutrino-antineutrino pair emission (positive slope) and $\delta t_\mathrm{LIV}^{(1)}(E=100\,\mathrm{TeV})$ (negative slope) as a function of $\Lambda/M_P$ and for $z=0.1$ (solid), 1 (dashed) and 3 (dotted lines).}
    \label{fig:EmaxDt_n1}
\end{figure}

The conclusion that the maximum energy of superluminal particles is below the energy range of the observed high energy astrophysical neutrinos still holds for a wide range of greater values of the LIV scale, thanks to the dependence on $\Lambda$ of the maximum energy, $E_\mathrm{max}^{(1)} \propto \Lambda^{3/8}$. At the same time, since $\delta^{(1)}_\mathrm{LIV} \propto \Lambda^{-1}$, the effect on the time of flight would become smaller as one increases the LIV scale. For example, observable time delays for a 100\,TeV astrophysical neutrino are incompatible with such neutrino being superluminal for any value of $\Lambda$, as Fig.~\ref{fig:EmaxDt_n1} shows.

\begin{figure}[tbp]
    \centering
    \includegraphics[height=8cm]{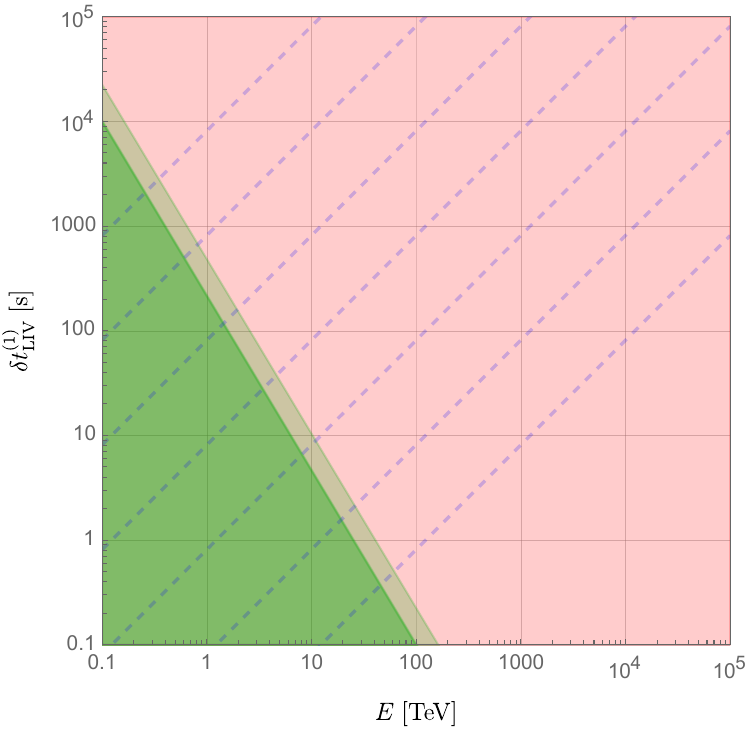}
    \caption{Allowed (green) and excluded (red) regions for superluminal neutrino events, for $n=1$, and from the lightest to the darker green, for $z=0.1$, 1 and 3 (let us note that in this graph the cases $z=1$ and 3 are visually indistinguishable). Dashed blue lines show points with constant $\Lambda$ (from top-left to bottom-right: $\log_{10}(\Lambda/M_P)=-2,-1,0,1,2,3$ and $4$). Compatible superluminal neutrino events should align along a single line, corresponding to a definite value of $\Lambda$, and be contained inside the green region.}
    \label{fig:exclusion_n1}
\end{figure}

An alternative way to present the correlation between the effect of LIV in the time of flight and in the energy spectrum of astrophysical neutrinos is based on the identification of the range of values of $(E,\delta t)$ consistent with LIV. This produces an ``exclusion'' diagram of allowed and forbidden regions, as in the log-log plot of Fig.~\ref{fig:exclusion_n1}. For a given energy $E$, the value of $\delta t$ would increase for a lower LIV scale, but this would also favour neutrino decays, so that the expected value of $\delta t$ may be incompatible with the observation of such energy (red region). The compatibility between a time delay $\delta t$ for a neutrino of energy $E$ and the observation of such neutrino defines the green region in this figure. All the early events should lay in the interior of this region, over a line with slope one, corresponding to a certain value of the LIV scale. However, Fig.~\ref{fig:exclusion_n1} shows that observable time delays of superluminal neutrinos for $n=1$ (green region in this figure) require energies of astrophysical neutrinos that lie in the region dominated by the atmospheric neutrino background.

Another consequence of the identification of a maximum energy in early events is that, accepting the interpretation of the highest energy neutrino event seen by IceCube~\cite{IceCube:2021rpz} as due to the resonant production of a $W$ by an electron-antineutrino interacting with an atomic electron in ice [the Glashow resonance~\cite{Glashow:1960zz}, peaking at an antineutrino energy of 6.3 petalectronvolts (PeV) in the rest frame of the electron], and assuming a linear LIV such that $\Lambda\lesssim 10^{8} M_P$ (so that $E_\mathrm{max}^{(1)}<6.3\,\mathrm{PeV}$), antineutrinos would be subluminal and neutrinos superluminal. To test this conclusion, however, one should be able to discriminate between neutrinos and antineutrinos, which is not possible with the present detection techniques.

\begin{figure}[tbp]
    \centering
    \includegraphics[height=8cm]{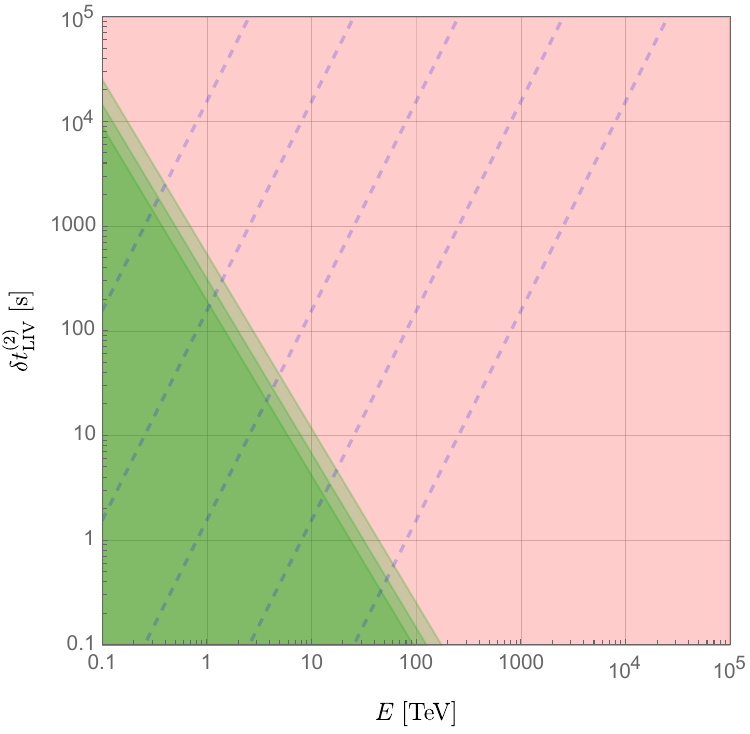}
    \caption{Allowed (green) and excluded (red) regions for superluminal neutrino events, for $n=2$, and from the lightest to the darker green, for $z=0.1$, 1 and 3. Dashed blue lines show points with constant $\Lambda$ (from top-left to bottom-right: $\log_{10}(\Lambda/M_P)=-9,-8,-7,-6$ and $-5$). Compatible superluminal neutrino events should align along a single line, corresponding to a definite value of $\Lambda$, and be contained inside the green region.}
    \label{fig:exclusion_n2}
\end{figure}

Another scenario, which is simpler to analyse, is the quadratic ($n=2$) purely superluminal\footnote{In the purely subluminal case there are no decays and then the time of flight is the only observable effect of LIV.} LIV neutrino model. According to Eq.~(\ref{Emax2}), the observation of astrophysical neutrinos above a few PeV ($E_\mathrm{max}^{(2)}>6.3\,\mathrm{PeV}$) implies that one should have a scale of LIV $\Lambda\gtrsim 10^{-2}\,M_P$. However, the effect of LIV in the time of flight for such a scale becomes unobservable (see Fig.~\ref{fig:exclusion_n2}).

In fact, such scenario was considered in previous works~\cite{Stecker:2014oxa,Carmona:2019xxp} as a possible explanation of the indications, prior to the discovery of a neutrino event compatible with the Glashow resonance, of a cutoff in the diffuse astrophysical neutrino spectrum extracted from IceCube data.
Future experiments (IceCube-Gen2, KM3NeT, and other instruments~\cite{Stecker:2022tzd}) will be able to measure the neutrino spectrum up to energies of the order of EeV ($10^{18}\,$eV). An unexpected cutoff in those observations would be a signal of the purely superluminal neutrino LIV scenario.

\section{Effects on the local propagation of neutrinos and antineutrinos}
\label{local}

One may wonder whether, in the case of non-astrophysical neutrinos, it would be possible to observe the decay effects of superluminal neutrinos or put better bounds on the LIV scale $\Lambda$.

When one considers the propagation of a superluminal particle that is produced at a distance $L$ small enough to neglect the effects of the expansion of the universe, it is convenient to define an energy $E_r^{(n)}(L)$ as the solution of 
\be
    \Gamma(E)|_{E=E_r^{(n)}(L)} \,=\, 1/L\,,
\ee
instead of Eq.~(\ref{Er(z)}). When the superluminal particle is produced with an energy $E>E_r^{(n)}(L)$ it will decay in a distance much smaller than $L$, while in the case $E<E_r^{(n)}(L)$ it would need a distance much larger than $L$ to decay, so that it will be detected before decaying. Once more, one can estimate that all the superluminal particles will be detected with an energy $E_d$
\be
    E_d < E^{(n)}_\mathrm{max}(L) \approx E_r^{(n)}(L)\,.
\ee

If we consider again values of the scale of new physics such that the neutrino-antineutrino emission determines the cutoff, $E^{(n)}_\mathrm{max}(L)$ is
\be 
    E^{(1)}_\mathrm{max}(L) \approx 3.01 \times 10^{3}\,\mathrm{TeV} \,\left( \frac{L}{\mathrm{km}} \right)^{-1/8} \left( \frac{\Lambda}{M_P} \right)^{3/8}
    \label{eq:er1L}
\ee
for $n=1$, and 
\be 
    E^{(2)}_\mathrm{max}(L) \approx 8.27\times 10^{6} \,\mathrm{TeV} \,\left( \frac{L}{\mathrm{km}} \right)^{-1/11} \left( \frac{\Lambda}{M_P} \right)^{6/11}
    \label{eq:er2L}
\ee
for $n=2$.

A comparison of Eqs.~(\ref{Emax1})-(\ref{Emax2}) with Eqs.~(\ref{eq:er1L})-(\ref{eq:er2L}) shows that the effect of LIV on the spectrum of superluminal particles for astrophysical neutrinos is stronger than for atmospheric neutrinos, as the maximum energy (cutoff) is several orders of magnitude smaller in the case of astrophysical neutrinos. 

One would therefore expect to find clearer signals of these anomalous decays in the study of astrophysical neutrinos. This is certainly the case for $n=2$ since, while we have detected astrophysical neutrinos of energies close to the PeV scale, which is the scale of the cutoff for $(\Lambda/M_P) \sim 10^{-2}$,  the corresponding local cutoff of $10^6\,$TeV is far above the highest energies of detected atmospheric neutrinos.

The case $n=1$, however, might be different, since, although the typical scale for the cutoff in the case of astrophysical neutrinos is rather low (of the order of the TeV or lower), if neutrinos were superluminal, then antineutrinos would be subluminal, and we could be receiving an astrophysical flux that would correspond only to antineutrinos. Such an hypothesis is consistent with the observation of the Glashow resonance and the time-of-flight-anomalous results found in~\cite{Amelino-Camelia:2022pja} and discussed in the previous section.

The question, then, is whether one could confirm or discard this scenario with the experimental constraints we have from atmospheric neutrino physics. Using the value of $(\Lambda/M_P)\sim(1/10)$ considered in the previous section, we have that $E_\mathrm{max}^{(1)}(10\,\mathrm{km}) \sim 950\,\mathrm{TeV}$ (corresponding to the typical distance traversed by down-going neutrinos), and $E_\mathrm{max}^{(1)}(10^4\,\mathrm{km}) \sim 350\,\mathrm{TeV}$ (corresponding to the typical distance traversed by up-going neutrinos).

The previous scenario would therefore mean a suppression of the atmospheric neutrino flux (neutrinos would disappear, and antineutrinos would not be affected) and a change in its standard flavour composition (originated from neutrino splitting) at energies which are rather close to the end of the atmospheric neutrino spectrum, which would certainly complicate the analysis. Nevertheless, it might be interesting to study the sensitivity of the data to the value of the scale $\Lambda$ and investigate the constraints that atmospheric neutrino physics is able to put on such an LIV scenario.

\section{Conclusions}
\label{conclusions}

There has recently been growing interest in studying LIV effects in the neutrino sector that result in a time delay between neutrinos and photons originating from astrophysical sources. These studies are motivated by both theoretical and experimental considerations. On the one hand, they serve as a valuable testing ground for probing quantum-gravity models, which typically predict modifications to the kinematics of standard relativity at extremely high energies.  On the other hand, the exploration of time-of-flight anomalies may provide insights into the puzzling absence of neutrino detections in correlation with gamma-ray burst (GRB) events~\cite{ANTARES:2013ath,ANTARES:2020vzs,IceCube:2022rlk}. This absence may signify a gap in our understanding, either in the mechanisms of neutrino production at these sources or in their journey to Earth. 

However, focusing solely on the temporal and directional correlations between neutrino events and GRBs for the investigation of possible LIV effects may not be enough to ensure a comprehensive and consistent analysis of the phenomenon. In the present work, we have demonstrated the necessity of concurrently considering LIV effects on the spectrum of superluminal particles, in conjunction with those affecting the time of flight due to the modified energy dependence of the velocity of propagation.

Specifically, we have shown that, excluding values of the LIV scale larger than the Planck scale, the dominant decay channel of superluminal particles is the neutrino splitting (production of a neutrino-antineutrino pair) of a parent neutrino. We have then given quantitative estimates for the maximum detected energy of astrophysical neutrinos and compared the sensitivity of such a cutoff in the spectrum with the magnitude of time delays between neutrinos and photons in different scenarios. 

In the case of a linear model of LIV, we have used the estimate for the cutoff in the neutrino spectrum to exclude the possibility to have observable early GRB-neutrino events (see Fig.~\ref{fig:exclusion_n1}). In the quadratic case, for a purely superluminal model, it is possible to use the observation of neutrino events at the highest energies to obtain bounds on the LIV scale that lead to conclude that there cannot be observable effects on time of flights (Fig.~\ref{fig:exclusion_n2}).

The modified neutrino kinematics is also present in the propagation of non-astrophysical neutrinos. However, the effects of a Lorentz invariance violation are stronger for astrophysical neutrinos, as they are amplified by the propagation over cosmological distances. This is consistent with our estimation of the maximum detected energy for atmospheric neutrinos in the quadratic case, which, for a scale $(\Lambda/M_P)\sim (1/10)$, is far above 100 TeV, the characteristic scale at which astrophysical neutrinos start to dominate over atmospheric neutrinos. 
In the linear case, however, the cutoff appears at energies of the order of just hundreds of TeV. Therefore, it might be interesting to investigate the consequences of this scenario in the atmospheric neutrino spectrum.

Let us finally note that the present work assumes that LIV occurs within the framework of effective field theory; however, it is possible to evade the constraints on LIV from the decays of superluminal particles if one goes beyond this framework. One example of such a scenario is a deformation of the Lorentz transformations that is compatible with the relativistic principle (DSR)~\cite{Amelino-Camelia:2000stu,Amelino-Camelia:2000cpa}. This deformation prevents the decay of massless superluminal particles. In addition, however, the description of time delays is much more involved than in the LIV case  and they might even not exist at all~\cite{Amelino-Camelia:2011ebd, Rosati:2015pga,Carmona:2017oit,Carmona:2022pro}. Other scenarios that can also evade the mentioned constraints are some based on string theory, such as the D-brane foam model considered in Ref.~\cite{Li:2023wlo}, which leads to a modification of the kinematics of superluminal particle decays.




\ack

We thank Antoine Kouchner for useful discussions during his visit to University of Zaragoza on experimental studies about the search of correlations between astrophysical neutrinos and GRBs. This work is supported by the Spanish grants PGC2022-126078NB-C21, funded by MCIN/AEI/ 10.13039/501100011033 and `ERDF A way of making Europe’, grant E21\_23R funded by the Aragon Government and the European Union, and the NextGenerationEU Recovery and Resilience Program on `Astrofísica y Física de Altas Energías’ CEFCA-CAPA-ITAINNOVA. The work of M.A.R. is supported by the FPI grant PRE2019-089024, funded by MICIU/AEI/FSE. The authors would like to acknowledge the contribution of the COST Action CA18108 ``Quantum gravity phenomenology in the multi-messenger approach''.

\section*{References}

\bibliographystyle{JHEP}
\bibliography{bib/QuGraPheno.bib}

\providecommand{\href}[2]{#2}\begingroup\raggedright\begin{thebibliography}{10}

\bibitem{Colladay:1998fq}
D.~Colladay and V.A.~Kostelecky, \emph{{Lorentz violating extension of the
  standard model}},
  \href{https://doi.org/10.1103/PhysRevD.58.116002}{\emph{Phys. Rev.}
  {\bfseries D58} (1998) 116002}
  [\href{https://arxiv.org/abs/hep-ph/9809521}{{\ttfamily hep-ph/9809521}}].

\bibitem{Carmona:2022lyg}
J.M.~Carmona, J.L.~Cort\'es, J.J.~Relancio and M.A.~Reyes, \emph{{Cosmic
  Neutrinos as a Window to Departures from Special Relativity}},
  \href{https://doi.org/10.3390/sym14071326}{\emph{Symmetry} {\bfseries 14}
  (2022) 1326} [\href{https://arxiv.org/abs/2206.14257}{{\ttfamily
  2206.14257}}].

\bibitem{Ellis:2002in}
J.R.~Ellis, N.~Mavromatos, D.V.~Nanopoulos and A.S.~Sakharov,
  \emph{{Quantum-gravity analysis of gamma-ray bursts using wavelets}},
  \href{https://doi.org/10.1051/0004-6361:20030263}{\emph{Astron. Astrophys.}
  {\bfseries 402} (2003) 409}
  [\href{https://arxiv.org/abs/astro-ph/0210124}{{\ttfamily
  astro-ph/0210124}}].

\bibitem{Jacob:2008bw}
U.~Jacob and T.~Piran, \emph{{Lorentz-violation-induced arrival delays of
  cosmological particles}},
  \href{https://doi.org/10.1088/1475-7516/2008/01/031}{\emph{JCAP} {\bfseries
  01} (2008) 031} [\href{https://arxiv.org/abs/0712.2170}{{\ttfamily
  0712.2170}}].

\bibitem{ANTARES:2016fmg}
{\scshape ANTARES} collaboration, \emph{{Stacked search for time shifted high
  energy neutrinos from gamma ray bursts with the ANTARES neutrino telescope}},
  \href{https://doi.org/10.1140/epjc/s10052-016-4496-8}{\emph{Eur. Phys. J. C}
  {\bfseries 77} (2017) 20} [\href{https://arxiv.org/abs/1608.08840}{{\ttfamily
  1608.08840}}].

\bibitem{Zhang:2022svg}
H.~Zhang and L.~Yang, \emph{{Testing Lorentz Violation with IceCube
  Neutrinos}}, \href{https://doi.org/10.3390/universe8050260}{\emph{Universe}
  {\bfseries 8} (2022) 260}.

\bibitem{Huang:2022xto}
Y.~Huang and B.-Q.~Ma, \emph{{Ultra-high energy cosmic neutrinos from gamma-ray
  bursts}},  \href{https://arxiv.org/abs/2211.00231}{{\ttfamily 2211.00231}}.

\bibitem{Amelino-Camelia:2022pja}
G.~Amelino-Camelia, M.G.~Di~Luca, G.~Gubitosi, G.~Rosati and G.~D'Amico,
  \emph{{Could quantum gravity slow down neutrinos?}},
  \href{https://doi.org/10.1038/s41550-023-01993-z}{\emph{Nature Astron.}
  {\bfseries 7} (2023) 996} [\href{https://arxiv.org/abs/2209.13726}{{\ttfamily
  2209.13726}}].

\bibitem{Carmona:2022dtp}
J.M.~Carmona, J.L.~Cort\'es, J.J.~Relancio and M.A.~Reyes, \emph{{Decay of
  superluminal neutrinos in the collinear approximation}},
  \href{https://doi.org/10.1103/PhysRevD.107.043001}{\emph{Phys. Rev. D}
  {\bfseries 107} (2023) 043001}
  [\href{https://arxiv.org/abs/2210.02222}{{\ttfamily 2210.02222}}].

\bibitem{Carmona:2012tp}
J.M.~Carmona, J.L.~Cortes and D.~Mazon, \emph{{Uncertainties in Constraints
  from Pair Production on Superluminal Neutrinos}},
  \href{https://doi.org/10.1103/PhysRevD.85.113001}{\emph{Phys. Rev. D}
  {\bfseries 85} (2012) 113001}
  [\href{https://arxiv.org/abs/1203.2585}{{\ttfamily 1203.2585}}].

\bibitem{Mattingly:2009jf}
D.M.~Mattingly, L.~Maccione, M.~Galaverni, S.~Liberati and G.~Sigl,
  \emph{{Possible cosmogenic neutrino constraints on Planck-scale Lorentz
  violation}}, \href{https://doi.org/10.1088/1475-7516/2010/02/007}{\emph{JCAP}
  {\bfseries 02} (2010) 007} [\href{https://arxiv.org/abs/0911.0521}{{\ttfamily
  0911.0521}}].

\bibitem{Stecker:2014oxa}
F.W.~Stecker, S.T.~Scully, S.~Liberati and D.~Mattingly, \emph{{Searching for
  Traces of Planck-Scale Physics with High Energy Neutrinos}},
  \href{https://doi.org/10.1103/PhysRevD.91.045009}{\emph{Phys. Rev.}
  {\bfseries D91} (2015) 045009}
  [\href{https://arxiv.org/abs/1411.5889}{{\ttfamily 1411.5889}}].

\bibitem{Reyes:2023osq}
M.A.~Reyes, D.~Boncioli, J.M.~Carmona and J.L.~Cort\'es, \emph{{Testing Lorentz
  invariance violation using cosmogenic neutrinos}},
  \href{https://doi.org/10.22323/1.444.1026}{\emph{PoS} {\bfseries ICRC2023}
  (2023) 1026} [\href{https://arxiv.org/abs/2309.02103}{{\ttfamily
  2309.02103}}].

\bibitem{Mattingly:2005re}
D.~Mattingly, \emph{{Modern tests of Lorentz invariance}}, {\emph{Living
  Rev.Rel.} {\bfseries 8} (2005) 5}
  [\href{https://arxiv.org/abs/gr-qc/0502097}{{\ttfamily gr-qc/0502097}}].

\bibitem{Stecker:2022tzd}
F.W.~Stecker, \emph{{Testing Lorentz Invariance with Neutrinos}},
  \href{https://arxiv.org/abs/2202.01183}{{\ttfamily 2202.01183}}.

\bibitem{Workman:2022ynf}
{\scshape Particle Data Group} collaboration, \emph{{Review of Particle
  Physics}}, \href{https://doi.org/10.1093/ptep/ptac097}{\emph{PTEP} {\bfseries
  2022} (2022) 083C01}.

\bibitem{ReyesHung:2023udr}
M.A.~Reyes, \emph{Exploration of Possible Signals beyond Special Relativity
  Using High-Energy Astroparticle Physics}, Ph.D. thesis, Universidad de
  Zaragoza, July, 2023.
\newblock \href{https://arxiv.org/abs/2307.03462}{{\ttfamily 2307.03462}}.

\bibitem{Amelino-Camelia:2016ohi}
G.~Amelino-Camelia, G.~D'Amico, G.~Rosati and N.~Loret,
  \emph{{In-vacuo-dispersion features for GRB neutrinos and photons}},
  \href{https://doi.org/10.1038/s41550-017-0139}{\emph{Nature Astron.}
  {\bfseries 1} (2017) 0139}
  [\href{https://arxiv.org/abs/1612.02765}{{\ttfamily 1612.02765}}].

\bibitem{IceCube:2021rpz}
{\scshape IceCube} collaboration, \emph{{Detection of a particle shower at the
  Glashow resonance with IceCube}},
  \href{https://doi.org/10.1038/s41586-021-03256-1}{\emph{Nature} {\bfseries
  591} (2021) 220} [\href{https://arxiv.org/abs/2110.15051}{{\ttfamily
  2110.15051}}].

\bibitem{Glashow:1960zz}
S.L.~Glashow, \emph{{Resonant Scattering of Antineutrinos}},
  \href{https://doi.org/10.1103/PhysRev.118.316}{\emph{Phys. Rev.} {\bfseries
  118} (1960) 316}.

\bibitem{Carmona:2019xxp}
J.M.~Carmona, J.L.~Cortes, J.J.~Relancio and M.A.~Reyes, \emph{{Lorentz
  violation footprints in the spectrum of high-energy cosmic neutrinos:
  Deformation of the spectrum of superluminal neutrinos from electron-positron
  pair production in vacuum}},
  \href{https://doi.org/10.3390/sym11111419}{\emph{Symmetry} {\bfseries 11}
  (2019) 1419} [\href{https://arxiv.org/abs/1911.12710}{{\ttfamily
  1911.12710}}].

\bibitem{ANTARES:2013ath}
{\scshape ANTARES} collaboration, \emph{{Search for muon neutrinos from
  gamma-ray bursts with the ANTARES neutrino telescope using 2008 to 2011
  data}}, \href{https://doi.org/10.1051/0004-6361/201322169}{\emph{Astron.
  Astrophys.} {\bfseries 559} (2013) A9}
  [\href{https://arxiv.org/abs/1307.0304}{{\ttfamily 1307.0304}}].

\bibitem{ANTARES:2020vzs}
{\scshape ANTARES} collaboration, \emph{{Constraining the contribution of
  Gamma-Ray Bursts to the high-energy diffuse neutrino flux with 10 yr of
  ANTARES data}}, \href{https://doi.org/10.1093/mnras/staa3503}{\emph{Mon. Not.
  Roy. Astron. Soc.} {\bfseries 500} (2020) 5614}
  [\href{https://arxiv.org/abs/2008.02127}{{\ttfamily 2008.02127}}].

\bibitem{IceCube:2022rlk}
{\scshape IceCube, Fermi Gamma-ray Burst Monitor} collaboration,
  \emph{{Searches for Neutrinos from Gamma-Ray Bursts Using the IceCube
  Neutrino Observatory}},
  \href{https://doi.org/10.3847/1538-4357/ac9785}{\emph{Astrophys. J.}
  {\bfseries 939} (2022) 116}
  [\href{https://arxiv.org/abs/2205.11410}{{\ttfamily 2205.11410}}].

\bibitem{Amelino-Camelia:2000stu}
G.~Amelino-Camelia, \emph{{Relativity in space-times with short distance
  structure governed by an observer independent (Planckian) length scale}},
  \href{https://doi.org/10.1142/S0218271802001330}{\emph{Int. J. Mod. Phys. D}
  {\bfseries 11} (2002) 35}
  [\href{https://arxiv.org/abs/gr-qc/0012051}{{\ttfamily gr-qc/0012051}}].

\bibitem{Amelino-Camelia:2000cpa}
G.~Amelino-Camelia, \emph{{Testable scenario for relativity with minimum
  length}}, \href{https://doi.org/10.1016/S0370-2693(01)00506-8}{\emph{Phys.
  Lett. B} {\bfseries 510} (2001) 255}
  [\href{https://arxiv.org/abs/hep-th/0012238}{{\ttfamily hep-th/0012238}}].

\bibitem{Amelino-Camelia:2011ebd}
G.~Amelino-Camelia, N.~Loret and G.~Rosati, \emph{{Speed of particles and a
  relativity of locality in $\kappa$-Minkowski quantum spacetime}},
  \href{https://doi.org/10.1016/j.physletb.2011.04.054}{\emph{Phys. Lett. B}
  {\bfseries 700} (2011) 150}
  [\href{https://arxiv.org/abs/1102.4637}{{\ttfamily 1102.4637}}].

\bibitem{Rosati:2015pga}
G.~Rosati, G.~Amelino-Camelia, A.~Marciano and M.~Matassa,
  \emph{{Planck-scale-modified dispersion relations in FRW spacetime}},
  \href{https://doi.org/10.1103/PhysRevD.92.124042}{\emph{Phys. Rev.}
  {\bfseries D92} (2015) 124042}
  [\href{https://arxiv.org/abs/1507.02056}{{\ttfamily 1507.02056}}].

\bibitem{Carmona:2017oit}
J.M.~Carmona, J.L.~Cortes and J.J.~Relancio, \emph{{Does a deformation of
  special relativity imply energy dependent photon time delays?}},
  \href{https://doi.org/10.1088/1361-6382/aa9ef8}{\emph{Class. Quant. Grav.}
  {\bfseries 35} (2018) 025014}
  [\href{https://arxiv.org/abs/1702.03669}{{\ttfamily 1702.03669}}].

\bibitem{Carmona:2022pro}
J.M.~Carmona, J.L.~Cort\'es, J.J.~Relancio and M.A.~Reyes, \emph{{Time delays,
  choice of energy-momentum variables, and relative locality in doubly special
  relativity}}, \href{https://doi.org/10.1103/PhysRevD.106.064045}{\emph{Phys.
  Rev. D} {\bfseries 106} (2022) 064045}
  [\href{https://arxiv.org/abs/2207.03799}{{\ttfamily 2207.03799}}].

\bibitem{Li:2023wlo}
C.~Li and B.-Q.~Ma, \emph{{Lorentz and CPT breaking in gamma-ray burst
  neutrinos from string theory}},
  \href{https://doi.org/10.1007/JHEP03(2023)230}{\emph{JHEP} {\bfseries 03}
  (2023) 230} [\href{https://arxiv.org/abs/2303.04765}{{\ttfamily
  2303.04765}}].

\end{thebibliography}\endgroup

\end{document}